\documentclass[conference]{IEEEtran}
\IEEEoverridecommandlockouts
\usepackage{cite}
\usepackage{amssymb,amsmath,amsthm, amsfonts}
\usepackage{algorithmic}
\usepackage{graphicx}
\usepackage{textcomp}
\usepackage{xcolor}
\newtheorem{theorem}{Theorem}

\usepackage[subrefformat=parens,labelformat=parens]{subfig} 

\usepackage{graphicx} 
\usepackage{hyperref} 
\usepackage{todonotes} 
\usepackage{breqn} 
\usepackage{tikz} 
\usepackage[shortlabels]{enumitem} 
\usepackage{array,multirow} 

\usepackage{amsmath}
\usepackage{amsfonts} 
\DeclareMathOperator*{\argmin}{arg\,min} 

\def\BibTeX{{\rm B\kern-.05em{\sc i\kern-.025em b}\kern-.08em
    T\kern-.1667em\lower.7ex\hbox{E}\kern-.125emX}}
\begin{document}

\title{How to Approximate any Objective Function via Quadratic Unconstrained Binary Optimization
}

\author{\IEEEauthorblockN{Thomas Gabor, Marian Lingsch Rosenfeld, Claudia Linnhoff-Popien}
\IEEEauthorblockA{\textit{Mobile and Distributed Systems} \\
\textit{LMU Munich}\\
Munich, Germany \\
thomas.gabor@ifi.lmu.de, m.rosenfeld@campus.lmu.de, linnhoff@ifi.lmu.de}
\and
\IEEEauthorblockN{Sebastian Feld}
\IEEEauthorblockA{\textit{Quantum and Computer Engineering} \\
\textit{Delft University of Technology}\\
Delft, Netherlands \\
s.feld@tudelft.nl}
}

\newcommand\copyrighttext{%
  \footnotesize This is a pre-print of an article published in \href{https://q-se.github.io/qsaner2022/}{Q-SANER 2022}.
  }
\newcommand\copyrightnotice{%
\begin{tikzpicture}[remember picture,overlay]
\node[anchor=south,yshift=10pt,xshift=10pt] at (current page.south) {\fbox{\parbox{\textwidth}{\copyrighttext}}};
\end{tikzpicture}%
}



\maketitle

\begin{abstract}
Quadratic unconstrained binary optimization (QUBO) has become the standard format for optimization using quantum computers, i.e., for both the quantum approximate optimization algorithm (QAOA) and quantum annealing (QA). We present a toolkit of methods to transform almost arbitrary problems to QUBO by (i) approximating them as a polynomial and then (ii) translating any polynomial to QUBO. We showcase the usage of our approaches on two example problems (ratio cut and logistic regression).
\end{abstract}

\copyrightnotice{}
\section{Introduction}
\label{sec:introduction}

Optimization using quantum computing is usually based on one of two major algorithms: the \textit{quantum approximate optimization algorithm} (QAOA)~\cite{farhi2014quantum} for gate-based quantum computers or \textit{quantum annealing} (QA)~\cite{kadowaki1998quantum} for adiabatic quantum computers. Both have in common that they can be applied directly to problems given in the form of \textit{quadratic unconstrained binary optimization} (QUBO). Subsequently, producing QUBO formulations for well-known to little-known problem classes has become an active part of research; examples of such applications originate from machine learning \cite{amin2018quantum, venturelli2019reverse, date2021qubo, willsch2020support}, scheduling \cite{ikeda2019application, venturelli2015quantum}, routing \cite{crispin2013quantum, feld2019hybrid, harwood2021formulating, syrichas2017large}, energy distribution \cite{silva2021qubo}, and many others.

Most authors consider a single application given via a single objective function \cite{calude2017qubo, neukart2017traffic, adachi2015application, silva2021qubo, dinneen2017formulating, yarkoni2021multi, souza2020application, titiloye2011quantum} while some provide approaches for several different problems \cite{jiang2018quantum, chang2019quantum, thom2021solving}. Finally, there are few overview articles in which best practices for modeling well-known theoretical problem classes are proposed \cite{lucas2014ising, glover2018tutorial}. However, problem modeling and problem transformation mostly remain non-generalizable, manual steps.


In this paper we focus on the generalization of the first transformation step. We present a list of methods to convert an arbitrary objective function into a polynomial (Sec. \ref{subsec:step1}). Then we describe a method to transform this polynomial in such a way that it will have the form of a quadratic unconstrained optimization problem (Sec. \ref{subsec:step2}). Thus, the original objective function can be minimized using QAOA or quantum annealing. We demonstrate the application and feasibility of our proposed methods using two separate problems, namely ratio cut and logistic regression (Sec. \ref{sec:evaluation}). Finally, we conclude with remarks on future work (Sec. \ref{sec:discussion}).
\section{Approaches}
\label{sec:concept}



Solving a QUBO problem can be defined as minimizing $x^T Q x$ for a symmetric matrix $Q\in \mathbb{R}^{n\times n}$ and $x\in\mathbb{B}^n$ returning
\begin{equation}
    \min_{x\in\mathbb{B}^n} \sum_{i=1}^{n}\sum_{j=1}^{n} Q_{ij} x_i x_j
\end{equation}


\textit{Quantum annealing} is an optimization method for QUBO problems using quantum phenomena \cite{kadowaki1998quantum}, whereas a physical implementation is provided by D-Wave Systems \cite{mcgeoch2014adiabatic, boothby2020next}. However, there also exist many classical solvers for QUBO problems~\cite{booth2017partitioning, aramon2019physics, streif2019comparison, van1987simulated}.






However, objective functions occurring ``in the wild'' may be of any structure, i.e., they can be represented by mathematical functions of any complexity. This also means that translating or approximating them as a QUBO formulation may become arbitrarily complex, both to the developer figuring out the approach and to the machine running a huge problem instance that is bloated from lots of translation overhead. The goal of this paper is to transform arbitrarily shaped objective functions into a QUBO representation as efficiently as possible. For this purpose we identify two main steps:

\begin{enumerate}[(A)]
    \item \textbf{Translate an objective function into a polynomial}
    \item \textbf{Translate the polynomial into QUBO form}
\end{enumerate}

For the better part of this paper, we present approaches that help developers follow through with these two steps in praxis and thus open up a new range of problem domains for quantum optimization.

\subsection{Translate an objective function into a polynomial}
\label{subsec:step1}

No general procedure can be given to create a polynomial or polynomial approximation from an arbitrary objective function. There are different approaches for special classes of functions, each with advantages and disadvantages. The usability depends, among other things, on the domain of the objective function to be transformed (e.g., $\mathbb{B}$ vs. $\mathbb{R}$ or $[-\infty, \infty]$ vs. $[-10,10]$) and also on the type (e.g., $ln(x)$ or $e^{x}$ are much easier to approximate than the cost function of an artificial neural network).

In the following we will present a collection of recommended actions, whereby the specific choice of action(s) --- and, if necessary, the order of application --- remains with the user. The list below serves as an overview and is sorted from more generic choices to more specific ones. In the upcoming subsubsections we will explain when and how each of these transformations should be applied.

\begin{enumerate}[({A}1)]
    \item \textbf{Approximate function via Lagrange polynomials}
    \item \textbf{Approximate function via splines}
    \item \textbf{Reduce complexity via additional constraints or changed function}
    \item \textbf{Apply bijective monotone functions}
    \item \textbf{Approximate periodic behavior via Fourier series}
    \item \textbf{Approximate differentiable function via Taylor series expansion}
\end{enumerate}



\subsubsection{Approximate function via Lagrange polynomials}\label{recipe:Lagrange}

Assume that we know the function values at some points of the otherwise unknown objective function. This can be the case, for example, for a given differential equation where the exact solution is not known, but that has been numerically approximated at some points.

Now Lagrange polynomials can be used to interpolate the objective function's known data points using polynomials. One advantage is that an arbitrary or unknown function can approximately be converted into a polynomial. For example, given points $(1, 2), (2, 6), (3, 12)$, a polynomial of degree $2$ extrapolating these points is $P(x) = 2\cdot\frac{(x - 2)(x - 3)}{(1 - 2)(1-3)} + 4\cdot\frac{(x - 1)(x - 3)}{(2 - 1)(2-3)} + 9\cdot\frac{(x - 1)(x - 2)}{(3 - 1)(3 - 2)} = x + x^2$. A huge disadvantage is that the degree of the polynomial used increases linearly with the number of data points involved in the interpolation. This means that this very generic approach should only be used if all other options fail.



Summary:
\begin{itemize}
    \item Use case: Objective function is unknown or very complex.
    \item Input: Few data points of the objective function.
    \item Output: Polynomial that approximates the objective function through given data points.
\end{itemize}

\subsubsection{Approximate function via splines}\label{recipe:splines}

Suppose that the given objective function is again unknown, only known in some intervals, or very complex. But this time Lagrange polynomials should not be used, because too many data points for interpolation and accordingly too many ancillary variables would be needed. Instead, we presume that the given objective function can be approximated at different intervals using multiple polynomials having low degree (usually degree $1$ or $3$) instead of one high-degree polynomial. Examples to this are stepwise defined functions or those with many data points known in some intervals.

In such cases splines can be used, i.e., functions that are piece-wise composed of polynomials. By means of such splines, polynomial approximations can be created for certain regions of a function, which are then joined together to obtain a continuous or even differentiable approximation of the original objective function. When using splines, however, it is important to ensure that they are limited to the objective function's exact intervals they are supposed to approximate.

An example shall clarify this: an objective function $f$ exists, which is to be interpolated in the intervals $I_1, \ldots , I_n$ (each containing a finite set of known data points) by polynomials $P_1, \ldots , P_n$. Now we can represent $f$ by the following approximated objective function $f'$ to be minimized:

\begin{equation}
\label{eq:splines}
	f'(x) = \sum_{i=1}^{n} (P_i (x)(\sum_{j\in I_i} y_j)) + \sum_{j \in I} y_j(j  - x )^2 + ( \sum_{j \in I} y_j - 1)^2
\end{equation}

\noindent with $I := \cup_{i=1}^{n} I_i$ being the union of intervals and $j$ are points of one particular interval or the union. Note that $j$s represent potential solutions and that they are dependent on the chosen accuracy of the intervals, i.e., they might be integers but also floating point numbers with two decimal places, for example. Furthermore, $y_j$ are binary slack variables that equal $1$ if $j=x$, otherwise $0$. To some extent, these slack variables create an if-then-else construct, as they are used to identify the interval the given point $x$ resides in.
The first term in Eq. \ref{eq:splines} effects that exactly that polynomial is evaluated which approximates the interval in which $x$ is located. 
If we consider the correct interval/polynom in which $x$ is located, then one of the slack variables $y_j$ is set to $1$, resulting in a multiplication with the value of the polynomial. If we are not in the correct interval, then the second sum of the first term is $0$ and so is the multiplication.
The remaining terms are needed to activate the correct $y_j$ (and only that), since depending on the given polynomials that are potentially defined on the whole domain, it might be the case that all or none will be chosen. For this, the second term effects that only the slack variable is activated for which $x=j$ and the third term assures that only a single $y_i$, and thus interval/polynomial, is chosen.
In a nutshell: all three constraints implement that the correct polynomial is taken for the considered value of $x$. 

Summary:
\begin{itemize}
    \item Use case: Objective function is unknown, defined in intervals or very complex.
    \item Input: Data points of objective function.
    \item Output: Function defined in intervals with each interval approximated by a polynomial based on the corresponding data points given.
\end{itemize}

\subsubsection{Reduce complexity via additional constraints or changed function}\label{recipe:constraint}

Sometimes it is worth considering whether the objective function has to be used in given form or whether it can be simplified without (major) loss of quality. There are different use cases where this might be possible.

First, checks that would otherwise be made after an optimization may be coded into the objective function as an additional explicit constraint. An example could be the requirement that at least one of the function's variables $x,y,z$ must be equal to $0$. In this case, instead of a subsequent check, the term $(xyz)^2$ can be added to the objective function to be minimized, which always imposes a penalty value unless at least one of the variables is equal to $0$, as desired.

Another possibility is at hand if a condition is implicitly contained in a given objective function. Then there is the opportunity that adding an explicit constraint might simplify the function. An example of this may be objective function $f(x,y,z)=10(x-y)^2+zx+zy$ with $x,y,z \in [-10, 10] \cap \mathbb{Z}$ that we want to minimize. It can be seen that the first term $10(x-y)^2$ will take its minimum of $0$ if $x=y$ holds. Since $x\neq y$ makes the cost function grow much more than the effect that $zx+zy=z(x+y)$ has, we can apply the explicit condition $x=y$. Doing this simplifies function $f$ to $f'(x,z)=2zx$.

Finally, various modifications can be made to the objective function itself to reduce its complexity and thus to improve the possibility of approximation using polynomials. These transformations include reducing discontinuities (e.g., replacing a step function by a sigmoid function), replacing an iterative optimization procedure by a single objective function (e.g., in the context of least square optimization), or changing the metric used (e.g., using $l^2$ metric instead of $l^4$ metric in the context of activation functions in artificial neural networks).

Summary:
\begin{itemize}
    \item Use case: Function contains implicit or explicit constraints or other possibilities for simplification.
    \item Input: Complex objective function, possibly with implicit constraints.
    \item Output: Simplified objective function.
\end{itemize}

\subsubsection{Apply bijective Monotone Functions}\label{recipe:monotone}

Bijective monotone increasing or decreasing functions have a special property: if applied to another function, the second function changes, but not the argument of its global optimum (if it exists). The derivation of this is quickly sketched using the example of a function to be minimized: Let $f:X \rightarrow \mathbb{R}$ be a minimization function and let $g:\mathbb{R} \rightarrow \mathbb{R}$ be a bijective monotone increasing function. If $x^* = \argmin_{x\in X} f(x)$, then $ \forall x\in X:f(x^*) \leq f(x) \iff \forall x\in X: g(f(x^*)) \leq g(f(x)) $ applies, because $g$ is both bijective and monotone increasing. In turn, this means that $\argmin_{x\in X} f(x) = \argmin_{x\in X}g(f(x))$, because $x^*$ is the minimal argument for both sides of the equation.

Two examples of bijective monotone functions are $e^x$ and $ln(x)$. They can be used to transform a multiplication into an addition and vice versa without changing the argument of the global optimum. Assume a polynomial optimization function of degree $4$, namely $f(x) = x^4$. If we now apply the natural logarithm, we receive $ln(f(x)) = 4 \cdot ln(x)$, thus a linear function with identical argument of the global minimum ($x=0$). Another example is objective function $\frac{f(x)}{g(x)}$. If we apply the natural logarithm again, we get $ln(f(x)) - ln(g(x))$, which can then be converted into a polynomial using Taylor series expansion (see subsection \ref{recipe:taylor}), as an example.


Summary:
\begin{itemize}
    \item Use case: Objective function has high degree, contains division/multiplication or other possibilities to be simplified by further function.
    \item Input: Complex objective function.
    \item Output: Simplified objective function.
\end{itemize}

\subsubsection{Approximate periodic behavior via Fourier Series}\label{recipe:fourier}

With Fourier series there is a possibility to approximate a periodic function by a sum of sine and cosine functions. Thus, complex periodic functions can be reduced into simpler periodic functions, which can then be approximated by polynomials in the interval of a single period.

\begin{figure}
\centering
	\includegraphics[width=.75\linewidth]{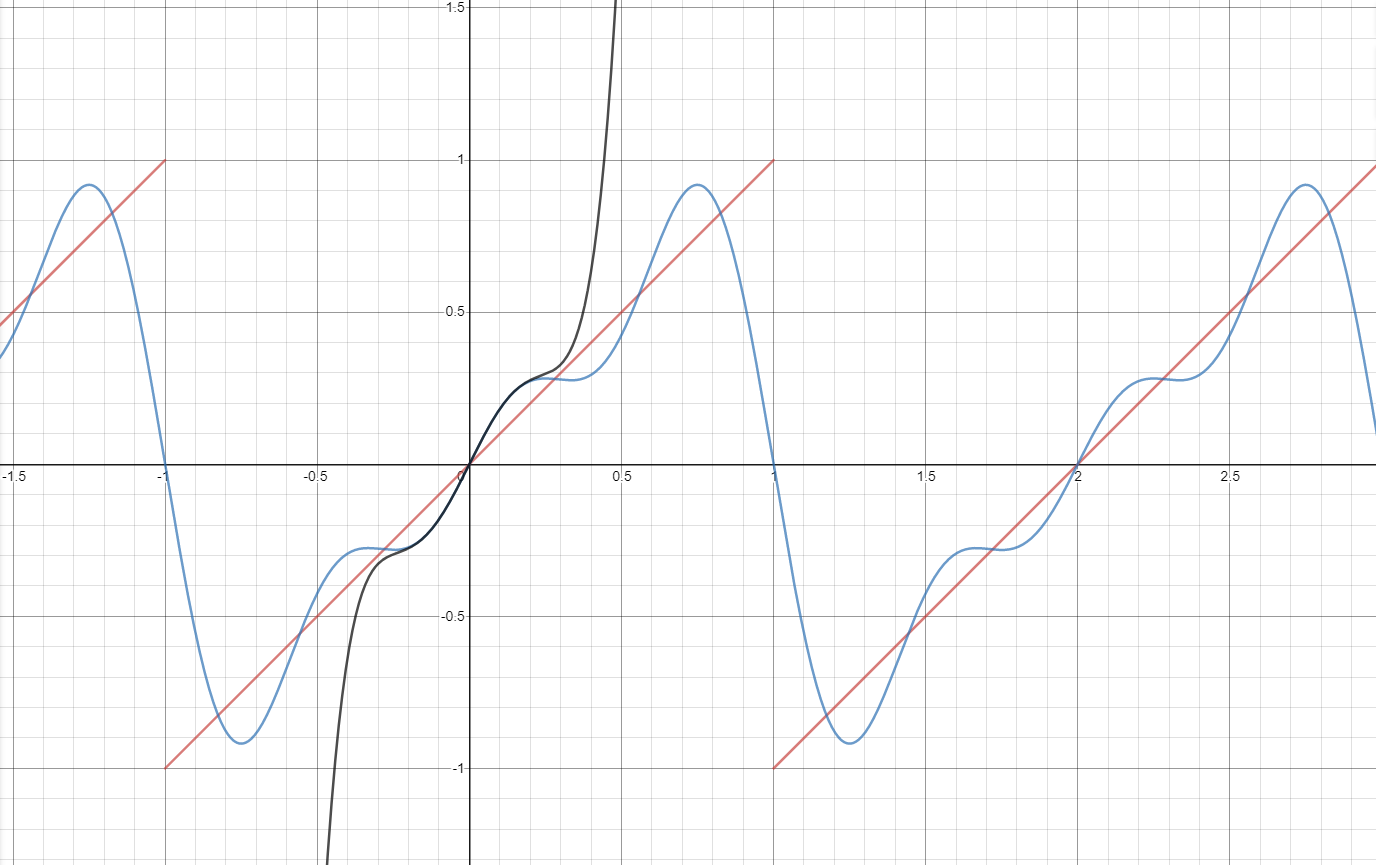}
    \caption{Approximation of periodic behavior using Fourier series. Red: exemplary objective function. Blue: approximation using Fourier series. Black: further approximation using Taylor series expansion.}
    \label{fig:fourier-series}
\end{figure}

As an example, objective function $f(x) = ((x-1) \text{ mod } 2) - 1$ is given (see Fig. \ref{fig:fourier-series}, red line). Now this function can be approximated by sine and cosine functions using Fourier series resulting in $f'(x) = \sum_{n=1}^{+\infty}2\cdot\frac{\left(-1\right)^{1+n}\cdot\sin\left(\pi nx\right)}{\pi n}$. With respect to a accuracy-complexity trade-off, we just utilized the first three terms (see Fig. \ref{fig:fourier-series}, blue line). Afterwards, this term can be converted into polynomials using Taylor series (see subsection \ref{recipe:taylor}). In our example, we performed a Taylor series expansion on the sinus functions of degree $5$ (see Fig. \ref{fig:fourier-series}, black line).
Putting these steps together, the approximated function is now $f''(x) = \sum_{n=1}^{3}2\cdot\frac{\left(-1\right)^{1+n}\cdot\left(\pi nx-\frac{\left(\pi n\right)^{3}}{6}x^{3}+\frac{\left(\pi n\right)^{5}}{120}x^{5}\right)}{\pi n}$.
Note that the performed steps (and in particular the Taylor series expansion) provide a good approximation only locally; one needs to apply these transformations carefully.

Summary:
\begin{itemize}
    \item Use case: Periodic objective function that cannot be approximated by Taylor series.\footnote{Note: If it can be approximated using Taylor series (e.g., $sin(x)$ or $cos(x)$), then this should be preferred to using Fourier series.}
    \item Input: Periodic objective function.
    \item Output: Objective function consisting of sum of sine and cosine functions.
\end{itemize}

\subsubsection{Approximate differentiable function via Taylor series expansion}\label{recipe:taylor}

\begin{figure}
\centering
	\includegraphics[width=.75\linewidth]{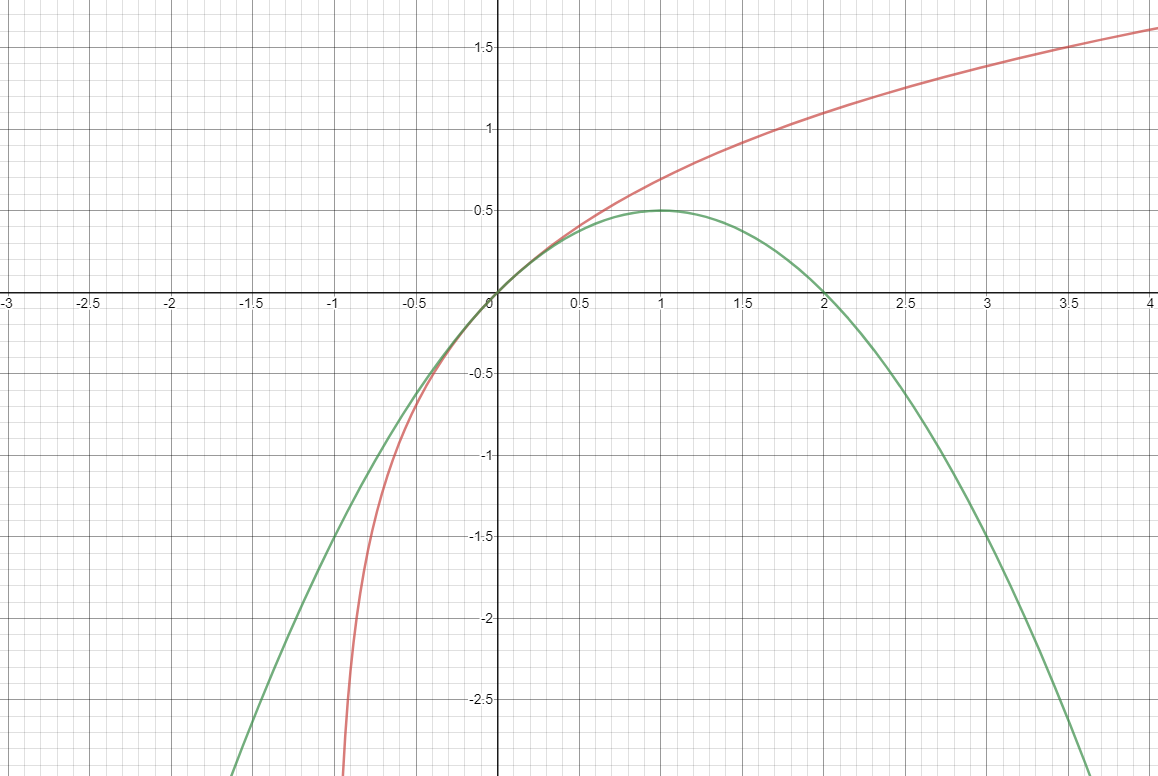}
    \caption{Approximation of the natural logarithm ($ln(x+1)$, red) using Taylor series ($x-\frac{x^2}{2}$, green). Note that the approximation is only locally good, see range $[-0.5,0.5]$.}
    \label{fig:log-approximation}
\end{figure}

If the given objective function is completely or partially differentiable, then the differentiable parts can be transformed into polynomials using Taylor series approximation. In doing so, however, there is a risk of introducing a potentially large error. Taylor series truncated after a finite set of terms, however, provide a good local approximation (see $ln(x+1) \approx x-\frac{x^2}{2}$ in Fig. \ref{fig:log-approximation}, here in particular region $[-0.5,0.5]$).

Thus, there is a trade-off between the approximation's quality and the number of qubits eventually needed, because more terms in the Taylor series lead to higher precision, which in turn results in a polynomial of higher degree and finally in more qubits required. Since, as described earlier, Taylor series provide only locally good approximations of a function, there should be a test whether the region in which the optimization is performed lies in the interval for which the approximation is of high quality. If this is not the case, a modification can be performed by addition or multiplication with constants in order to match search range with approximation interval. For example, if the Taylor approximation provides a good approximation for interval $[0,1]$, but the given objective function is defined on $[0,10]$, then scaling may help. Note, however, that this does not apply to all functions, as it may shift the function values. In Fig.~\ref{fig:log-approximation}, for example, the global maximum was shifted because the approximation has not been constrained to $[-0.5,0.5]$.

As another example for the approximation by means of Taylor series, consider an objective function that involves a division: $f(x,y) = \frac{x}{y}$. Applying the natural logarithm results in $ln(f(x,y))=ln(x)-ln(y)$. If we now approximate the natural logarithm using Taylor series up to order $2$, then $ln(z) = ln((z-1) + 1) \approx (z-1) - \frac{(z-1)^2}{2}$ follows. This in turn can be used to approximate objective function $f(x,y)$ as $f'(x,y) = (x-1) - \frac{(x-1)^2}{2} - (y-1) + \frac{(y-1)^2}{2}$, while the minimum's argument does not change (in range $[-0.5,0.5]$).

Summary:
\begin{itemize}
    \item Use case: Completely or partially differentiable objective function.
    \item Input: Differentiable objective function.
    \item Output: Polynomial objective function.
\end{itemize}

\subsection{Translate the polynomial into QUBO form}
\label{subsec:step2}

Having generated a polynomial $P:\mathbb{R}^n \rightarrow \mathbb{R}$ from our original problem formulation, we now need to translate that polynomial to a polynomial $P''':\mathbb{B}^{n'} \rightarrow \mathbb{R}$ that (as given by its signature) only uses binary variables and only uses them within quadratic terms. Note that during the translation process, we usually end up with $n' > n$ but, of course, try to keep the resulting overhead minimal. As polynomials have no means of containing hard constraints anyway, this polynomial $P'''$ is then in \emph{QUBO form}. For the construction we are going to present, Theorem~1 holds; however, we omit the proof here and instead refer the interested reader to the appendix.

\begin{theorem}\label{thm:correctness}
Let $P$ be a polynomial and $P'''$ the corresponding QUBO form produced according to the procedure we describe in Section II.B, then it holds that 
\begin{equation}\label{eq:theorem-minima}
\min_{x\in L} P(x_1, ..., x_n) = \min_{z\in \mathbb{B}^{n'}} P'''(z_1, ..., z_{n'})
\end{equation}
where $L = \{(x_1, ..., x_n) \; | \;  x_i \in \Lambda_i\}$ is the set of all possible solution candidates of $P$. Furthermore, there is a function $f: \mathbb{B}^{n'} \to L$  such that
\begin{equation}\label{eq:theorem-function}
\forall z^{*} \in \argmin_{z\in \mathbb{B}^{n'}}  P'''(z_1, ..., z_{n'}) : f(z^{*}) \in \argmin_{x\in L} P(x_1, ..., x_n).
\end{equation}
\end{theorem}


The intuition behind this step is as follows: As we consider the minimization of a polynomial $P:\mathbb{R}^n \rightarrow \mathbb{R}$ using QUBO, in the first step we translate each continuous variable in $P$ into a weighted sum of binary variables. In the second step, we introduce new variables to replace products of these binary variables. By doing this, we reduce the degree of multiplications and thus obtain the required quadratic form. In this case, these steps need to be applied in order:

\begin{enumerate}[({B}1)]
    \item \textbf{Re-scope to binary variables}
    \item \textbf{Restrict to quadratic terms}
\end{enumerate}

\subsubsection{Re-scope to binary variables}

The currently available quantum annealing hardware is still very limited, so the polynomial $P:\mathbb{R}^n \rightarrow \mathbb{R}$ should be represented as efficiently as possible, i.e., using as few decision variables as possible. Therefore, we restrict ourselves to minimizing the problem only with respect to a finite number of points, i.e., we reduce the search space for finding the minimum from $\mathbb{R}^n = \mathbb{R} \times \mathbb{R} \times \cdots \mathbb{R}$ to $\Lambda_1 \times \Lambda_2 \times \cdots \times \Lambda_n$, where each $\Lambda_i \subset \mathbb{R}$ and $\lvert \Lambda_i \rvert \in \mathbb{N}$. Thus, each of the polynomial's $n$ variables $x_i$ can only take values from the corresponding countable and finite set $\Lambda_i$. For example, a polynomial with $n=2$ variables might have domains $\Lambda_1=\{2,4,6\}$ with $\lvert \Lambda_1 \rvert=3$ and $\Lambda_2=\{-2,-1,0,1,2\}$ with $\lvert \Lambda_2 \rvert=5$.

Now the question arises how to minimally map elements $k$ of sets $\Lambda_i$ to corresponding binary decision variables $q_{i,k}$. A naive approach would be to represent each element by exactly one variable; thus $x_i = \sum_{k \in \Lambda_i} k \cdot q_{i,k}$ holds. With respect to the example above, we gain $x_1=6$ if $q_{1,3}=1$ since $x_1=2\cdot0+4\cdot0+6\cdot1=6$. However, it should be noted that a given value of a variable can also be composed as a sum of values, e.g., $6=2+4$. To avoid this, the term $\sum_{i}(\sum_{k \in \Lambda_i}q_{i,k} - 1)^2$ is added to the QUBO problem. This causes that exactly one value will be chosen, since otherwise a penalty term would be added. It is easy to see that this naive approach is quite inefficient, since one decision variable is needed for each variable value to be examined.

Instead of this one-hot encoding, a binary representation of the elements in $\Lambda_i$ can be used. By means of $x_i = \sum_{j=0}^{r_{\textit{max}}} 2^{j} \cdot q_{i,j}$ and $r_{\textit{max}}\in \mathbb{N}$ we can represent elements of $\lambda_i$ from $0$ up to a maximum number $2^{r_{\textit{max}}+1} - 1$. If, for example, $x_1=6$ is to be represented again, this is possible using $r_{\textit{max}}=2$ and $x_1 = 2^0 \cdot q_{1,0} + 2^1 \cdot q_{1,1} + 2^2 \cdot q_{1,2} = 2^0 \cdot 0 + 2^1 \cdot 1 + 2^2 \cdot 1 = 6$. Obviously, this representation is only suitable for positive natural numbers, whereas real numbers can only be approximated (the sum of natural numbers is always a natural number). To remedy this, we now represent variables as $x_i = \sum_{j=-r_{\textit{min}}}^{r_{\textit{max}}} 2^{j} \cdot q_{i,j}$ with $r_{\textit{min}}\in \mathbb{N}$. For example, value $6.5$ can be constructed as $2^{-1}+2^1+2^2$. Now, in order to also include negative numbers, another term is needed so that variables are finally represented by
\begin{equation}
    x_i = \sum_{j=-r_{\textit{min}}}^{r_{\textit{max}}} 2^{j} \cdot q_{i,j} - \sum_{j=-r_{\textit{min}}}^{r_{\textit{max}}} 2^{j} \cdot q_{i,j + (r_{\textit{max}} + r_{\textit{min}})}
\end{equation}
Although this term introduces some inefficiency as certain numbers may have multiple representations (e.g., $2^0=1$, but also $2^0+2^1-2^1=1$), it allows us to represent all numbers in $([-2^{r_{\textit{min}} + r_{\textit{max}} + 1} + 1, 2^{r_{\textit{min}} + r_{\textit{max}} + 1} - 1] \cap \mathbb{N}) \times 2^{-r_{\textit{min}}}$.

Abstractly formulated, we can now state that each variable $x_i$ no longer belongs to $\mathbb{R}$, but $\Lambda_i = \{\sum_{k=1}^{r}(B_i)_k \cdot t_{k} \; | \; t\in \{ 0,1 \}^{r}\}$, where we call $B_i$ the \emph{base}, with cardinality $r=|B_i|$, and $t$ the \emph{binary mask}. Note that with our example from above ($x_1 = 2^0 \cdot 0 + 2^1 \cdot 1 + 2^2 \cdot 1 = 6$), base $B_1$ would consist of $[2^0,2^1,2^2]$ with binary mask $t=[0,1,1]$. Consequently, each element of $\Lambda_i$ is represented as a weighted binary sum over the elements of $B_i$. Finally, the polynomial's variables are represented by $x_i = \sum_{b \in B_i} b \cdot q_{i,b}$ using elements $b$ of bases $B_i$ and binary masks represented by binary decision variables $q_{i,b}$.

In conclusion we consider polynomials of the form $P':\Lambda_1 \times \cdots \times \Lambda_n \rightarrow \mathbb{R}$ with degree $p \in \mathbb{N}$ and coefficients $a \in \mathbb{R}^{{(p+1)}^{n}}$ and represent them as follows:
\begin{equation}
    P'(x_1, ..., x_n) = \sum_{i_1, ..., i_n \in {0, 1, ..., p}} a_{i_1... i_n} \cdot x_1^{i_1}\cdots x_n^{i_n}
\end{equation}

An initial estimate on the number of binary decision variables needed in the following QUBO generation step is as follows: Given a polynomial with degree $p$, number of variables $n$, maximum multiplication degree $q$, and search range $r = r_{min} + r_{max}$, the value is upper bounded by $(n \cdot 2r)^{m}$ 
with $m = \lceil \frac{pq}{2} \rceil$. As an example, with $P'(x_1,x_2,x_3) = x_3^3 + x_1x_2 -1$ and $r=2+2=4$, it follows $p=3$, $n=3$, $q=2$, $m=3$, and consequently an upper bound of $(n \cdot 2r)^{m} = (3 \cdot 8)^{3} = 13824$ variables.
The worst case is thus given if $q$ variables to the $p$-th power are multiplied with each other in the polynomial (example: $x_1^3x_2^3$). However, this is not necessarily the case, which is why the number of binary decision variables required can be much lower. In the given case, the actual number is $2 \cdot 8 + 8^2 = 80$.

\subsubsection{Restrict to quadratic terms}\label{ssec:qubo-generation}

In the previous section, every (continuous) variable in $P$ was transformed into a sum of binary variables. Thus, from the original polynomial $P:\mathbb{R}^{n} \mapsto \mathbb{R}$ we derived polynomial $P':\mathbb{B}^{n'} \mapsto \mathbb{R}$ with $n' \in \mathbb{N}$. Now, $P'$ is brought into QUBO form, which means that we ensure that it only consists of at most square terms.

If no more than two binary variables are multiplied with each other in polynomial $P'$, a QUBO form is already given and nothing more needs to be done. If this is not the case, two new variables are introduced for every term in which more than two variables are multiplied with one another. These will each represent one of the two halves of the multiplied variables. For example, if 
\begin{equation}\label{ex:1}
    P'(x_1, x_2, x_3, x_4) = -x_1x_2x_3x_4 + x_4
\end{equation}
is given, two new variables $q_1, q_2$ with $q_1 = x_1x_2$ and $q_2 = x_3x_4$ are introduced. The previously cubic polynomial $P'$ is thus transformed into the quadratic polynomial:
\begin{equation}
P''(q_1, q_2, x_4) = -q_1q_2 + x_4    
\end{equation}
Note that it does not matter how we split up the original variables as long as we end up with $q_1 q_2$ as our new factor.

The global minimum of $P''$, however, is not necessarily identical to the global minimum of $P'$. This is based on the fact that, from the optimization function's point of view, the new variables $q_i$ have no relation to the original variables $x_j$. This can be seen in the example above, because with $q_1=1, q_2=1, x_4=0$ a global minimum is given for $P''$, but this constellation of variables constitutes a contradiction: $q_2=1$ implies $x_3=1$ and $x_4=1$, but we had set $x_4=0$. For this reason, the new variables $q_i$ must be connected back to the old variables $x_j$.

This is achieved by including the constraint 
\begin{equation}\label{eq:constraint}
q = x_1 \cdot x_2 \cdot ... \cdot x_n
\end{equation}
into the QUBO form to be minimized. In the following, we will explain that using the special case $q = q_1q_2$, which can be generalized by introducing variables $q_1 = x_1\cdots x_{\lfloor\frac{n}{2}\rfloor}$ and $q_2 = x_{\lfloor\frac{n}{2}\rfloor + 1}\cdots x_n$.
Constraint $q = x_1x_2$ is met iff penalty term 
\begin{equation}\label{eq:penalty-term}
4q -3qx_1 -3qx_2 + 2x_1x_2    
\end{equation}
has a global minimum of $0$. This term is added to the QUBO form and weighted with 
\begin{equation}\label{eq:A}
    A = 1 + \sum_{a \in P'}2|a|
\end{equation}
where $a \in P'$ denotes all the coefficients of polynomial $P'$. This weighting ensures that whatever value the polynomial takes when minimized, the constraint is always maintained and thus the introduction of the new variables remains consistent.

With respect to the example (see Eq. \ref{ex:1}), weighting $A = 1 + 2 + 2= 5$ follows for polynomial $P''$. For the two newly introduced variables we get penalty terms $4q_1 - 3q_1x_1 -3q_1x_2 + 2x_1x_2$ and $4q_2 - 3q_2x_3 -3q_2x_4 + 2x_3x_4$ and eventually the final polynomial in QUBO form:
\begin{multline}\label{ex:poly3}
P'''(x_1, x_2, x_3, x_4, q_1, q_2) = -q_1q_2 + x_4\\
+ 5(4q_1 -4q_1x_1 -3q_1x_2 + 2x_1x_2)\\
+ 5(4q_2 -4q_2x_3 -3q_2x_4 + 2x_3x_4)
\end{multline}
We see that $P'''$ (see Eq. \ref{ex:poly3}) is a polynomial of binary variables with a maximum of two variables being multiplied with each other.

\section{Examples}
\label{sec:evaluation}

We demonstrate the feasibility of our approach using two separate optimization problems.

\subsection{Ratio Cut}\label{ssec:ratio-cut}


In graph theory, a \textit{cut} denotes a partition of the set of the graph's vertices. In this example, we search for an optimal \emph{ratio cut}, i.e., given a graph $G = (V,E)$ with vertices $V$ and edges $E$ we search for two vertex sets $A, B \subset V$ with $A \cup B = V$ and $A \cap B = \emptyset$ so that

\begin{equation}\label{eq:n-cut-startkost}
rcut(A, B) = \frac{cut(A,B)}{|A|} + \frac{cut(A,B)}{|B|}
\end{equation}

\noindent becomes minimal. Here, $cut(A,B)$ denotes the number of edges between partitions $A$ and $B$. $|A|$ and $|B|$ indicate the number of vertices in $A$ and $B$, respectively.


Eq. \ref{eq:n-cut-startkost} clearly shows that it contains divisions of terms that cannot be represented as a polynomial. This is the case because the terms are not constant, but depend on variables $A,B$.

According to our proposal, the addition of constraints or modification of cost function (Sec. \ref{recipe:constraint}), the application of bijective monotonic functions (\ref{recipe:monotone}), or the realization of Taylor series expansion (\ref{recipe:taylor}) comes into question. In the given case we decide to modify the cost function (Sec. \ref{recipe:constraint}) by combining the fractions and obtaining a common denominator. Since $|A| + |B| = |V|$ and therefore constant, it follows:

\begin{equation}
rcut(A, B) \approx \frac{cut(A,B)(|A| + |B|)}{|A||B|} \approx \frac{cut(A,B)}{|A||B|}
\end{equation}

With this initial step we have received a single division, which will now be transformed into an addition using a logarithm (Sec. \ref{recipe:monotone}). Here we use the natural logarithm, since its Taylor expansion (Sec. \ref{recipe:taylor}) is easy to calculate. Thus, it follows:

\begin{equation}
rcut(A, B) \approx  \ln(cut(A,B)) - (\ln(|A|) + \ln(|B|))
\end{equation}

By using the natural logarithm, the division was replaced by a subtraction, which is why the next step to perform is a Taylor series expansion (Sec. \ref{recipe:taylor}). Using the constants $C_1, C_2, C_3, D_1, D_2, D_3 \in \mathbb{R}$, the approximation of $\ln$ can now be improved, because the following applies:

\begin{align}\label{eq:ln-approx}
\ln(x) &= \ln(C) + \ln\left(\frac{x}{C}\right) \\
&= \ln(C) + \ln\left(\left(\frac{x}{C} - D\right) + D\right) \\
&\approx \ln(C) + \ln(D) + \frac{\frac{x}{C} - D}{D} - \frac{(\frac{x}{C} - D)^2}{2 D^2} \\
&\approx \frac{\frac{x}{C} - D}{D} - \frac{(\frac{x}{C} - D)^2}{2 D^2}  
\end{align}

Note, that in last line of Eq. \ref{eq:ln-approx} the constant terms have been dropped, as they are irrelevant in the context of optimization. These constants influence the transformation of variable $x$ from the range $[x_0, x_1]$ into range $[\frac{x_0}{C} - D, \frac{x_1}{C} - D]$. Thus, $C_1, C_2, C_3, D_1, D_2, D_3 \in \mathbb{R}$ are parameters that can be changed in order to ``shift'' the range of $x$ to another range in which the Taylor expansion of the natural logarithm is (more) appropriate for our given application. The values of the constants can be determined either empirically or by analyzing the variable ranges and calculating the target ranges. The following example should clarify this: If the variables are of interest in the range $[-100, 200]$ and the corresponding Taylor approximation is appropriate in range $[0.5 , 1.5]$, then constants $C = 200, D = -1$ result in a meaningful transformation.

It follows:

\begin{align}
\begin{split}
rcut(A, B) \approx& \ln\left(\left(\frac{cut(A,B)}{C_1} - D_1\right) + D_1\right)\\
&- \ln\left(\left(\frac{|A|}{C_2} - D_2\right) + D_2\right)\\
&- \ln\left(\left(\frac{|B|}{C_3} - D_3\right) + D_3\right)
\end{split}
\end{align}

\begin{dmath}\label{eq:ncut-polynom}
rcut(A, B) \approx \frac{2cut(A,B)}{D_1C_1} - \frac{cut(A,B)^2}{2D_1^2C_1^2} - \frac{2|A|}{D_2C_2} + \frac{|A|^2}{2D_2^2C_2^2} \\ - \frac{2|B|}{D_3C_3} + \frac{|B|^2}{2D_3^2C_3^2}
\end{dmath}

Resulting Eq. \ref{eq:ncut-polynom} now shows that by using a Taylor expansion there is no longer a logarithm, no division by a variable takes place, and thus a polynomial for the original cost function $rcut(A,B)$ was found.

Now, $|A|$, $|B|$ and $cut(A, B)$ have to be represented using binary variables. To do this, we introduce variables $x_{i,j}$, where $x_{i,j} = 1$ means that vertex $i$ belongs to partition $j$. This results in the following translations:



\begin{equation}\label{eq:A-bin}
|A| = \sum_{n = 1}^{|V|} x_{n,0}, \;\;\;|B| = \sum_{n = 1}^{|V|} x_{n,1}
\end{equation}

\begin{equation}\label{eq:cut-bin}
cut(A,B) = \sum_{n,m = 1}^{|V|} x_{n,0}x_{m,1}
\end{equation}

Finally, an additional constraint is required (Sec. \ref{recipe:constraint}), which enforces that each vertex of the graph is assigned to exactly one partition. This is achieved with 

\begin{equation}\label{eq:cut-constraint-added}
A\sum_{n = 1}^{|V|} (x_{n,0} + x_{n,1} - 1)^2
\end{equation}

\noindent where weighting $A\in \mathbb{R}^{+}$ ensures that this constraint is prioritized within the minimization of the cost function.

If we now insert Eqs. \ref{eq:A-bin}, \ref{eq:cut-bin} into Eq. \ref{eq:ncut-polynom} and also add the constraint from Eq. \ref{eq:cut-constraint-added}, then we get a representation of the cost function from Eq. \ref{eq:n-cut-startkost} as a polynomial consisting only of binary variables. However, this polynomial is not yet in QUBO form as $cut(A, B)^2$ incorporates terms in which up to four binary variables are multiplied. Thus, the last necessary step is the application of the method presented in Sec. \ref{ssec:qubo-generation} in order to retrieve the final QUBO form. This step is not explicitly demonstrated here, since the resulting polynomial will hardly be readable.

\begin{figure}
\centering
\begin{tikzpicture}[node distance={15mm}, thick, main/.style = {draw, circle}] 
\node[main] (0) {$0$}; 
\node[main] (1) [above right of=0] {$1$}; 
\node[main] (2) [below right of=0] {$2$}; 
\node[main] (3) [above right of=2] {$3$}; 
\node[main] (4) [right of=3] {$4$}; 
\node[main] (5) [above right of=4] {$5$}; 
\node[main] (6) [below right of=4] {$6$}; 
\node[main] (7) [below right of=5] {$7$}; 
\draw (0) -- (1); 
\draw (0) -- (2); 
\draw (0) -- (3); 
\draw (1) -- (2); 
\draw (1) -- (3);
\draw (2) -- (3); 
\draw (3) -- (4); 
\draw (4) -- (5); 
\draw (4) -- (6); 
\draw (4) -- (7); 
\draw (5) -- (6);
\draw (5) -- (7);
\draw (6) -- (7);
\end{tikzpicture}
\caption{Graph used for demonstration. Partitions $A=[0,1,2,3]$ and $B=[4,5,6,7]$ were determined, which corresponds to the global optimum.}
\label{fig:graph-tikz}
\end{figure}
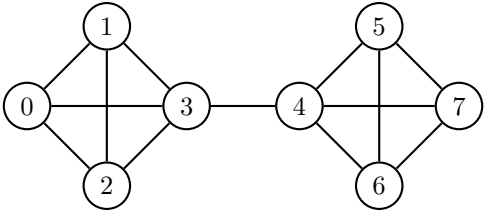

To demonstrate the general functionality of this whole translation, the graph from Fig. \ref{fig:graph-tikz} has been examined. The values of the constants, determined by analytical analysis followed by a heuristic search, are set to $D_1 = 1, C_1= 2, D_2 = 1, C_2 = 8, D_3= 1, C_3 = 8, A = 100$. Solving this particular problem instance returned partitions $A=[0,1,2,3]$ and $B=[4,5,6,7]$ as the global optimum.

\subsection{Logistic Regression}\label{ssec:logistic-regression}

\begin{figure}
\centering
   \includegraphics[width=.75\linewidth]{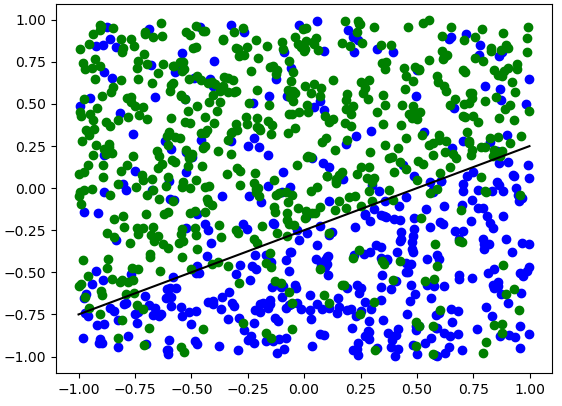}
   \caption{Example of a regression problem. Points with label $0$ shown in blue, points with label $1$ shown in green. The separating line is shown in black.}
    \label{fig:log-reg-example}
\end{figure}

The problem of logistic regression in $d$-dimensional space consists of a set of data points $(x_1, y_1), ..., (x_n, y_n), n \in \mathbb{N},$ where $x_i \in \mathbb{R}^d$ and $y_i \in \{0,1\}$ for all $i < n$. The goal is to find a function $f : \mathbb{R}^d \to \{0,1\}$ so that $y_i = f(x_i)$ holds for as many $x_i$ as possible, i.e., $f$ can predict the \emph{label} of an input $x_i$. Usually, there exist some tight constraints on the structure of $f$ so that simply looking up the labels in the data set is not feasible. Fig.~\ref{fig:log-reg-example} shows an example for $d=2$ and $f(x) = (mx > t)$ with parameters $m, t$, i.e., $f$ is a separating line in $2$-dimensional space.


We now translate a more general instance of logistic regression to QUBO as an additional showcase for our toolkit for QUBO translations. We choose a sigmoid separation function $\sigma_\theta(x) = \frac{1}{1 + e^{- \theta^T x}}$ so that $f(x) = (\sigma_\theta(x) > 0.5)$ for a single parameter vector $\theta \in \mathbb{R}^d$. 

We derive our optimization goal from the binary cross-entropy between the correctly and incorrectly classified sets:


\begin{equation}\label{eq:log-reg-cost-orig}
\sum_{i=1}^n -y_i \cdot \ln\left({\frac{1}{1+e^{-\theta^Tx_i}}}\right) - (1-y_i) \cdot \ln\left({1-\frac{1}{1+e^{-\theta^Tx_i}}}\right)
\end{equation}


Through application of Taylor series expansion (Sec.~\ref{recipe:constraint}) and simplification we can transform Eq.~\ref{eq:log-reg-cost-orig} into:




\begin{equation}
\sum_{i=1}^n \ln\left({1+e^{-\theta^Tx_i}}\right) + (1-y_i)\theta^Tx_i
\end{equation}

With Taylor expansion $\ln\left({1+e^{-x}}\right) \approx \ln{2} - \frac{1}{2}x + \frac{1}{8}x^2$ we derive:

\begin{equation}
 \sum_{i=1}^n (1-y_i)\theta^Tx_i -  \frac{1}{2}\theta^Tx_i + \frac{1}{8}(\theta^Tx_i)^2
\end{equation}

Note that it is easy to expand this example to more multidimensional labels, i.e., $y_i \in \{0,1\}^g$ for some $g \in \mathbb{N}$; all we need to do is use a corresponding matrix for $\theta$ instead of a vector.





To evaluate our method we produced a data set for linear regression in the following way: We sampled parameters $b,c$ for a function $f: \mathbb{R}^{d-1} \to \mathbb{R}$ with $f(x) = \sum_{i=0}^n x_{i} \cdot c_i + b$. 
This function separates $\mathbb{R}^d$ into to partitions, i.e., $\mathbb{R}^d = \{x \; | \; f(x) \leq 0 \} \; \cup \; \{x \; | \; f(x) > 0\}$. We also sampled $n$ data points $x_1, ..., x_n \in \mathbb{R}^d$. Each test run was defined by a probability $p \in [0;1]$ given as a parameter, which described how likely we were to assign --- to a data point $x_i$ --- a label $y_i$ that matches the separation performed by $f$: If $x_i \in \{x \; | \; f(x) \leq 0 \}$ we assign $y_i = 0$ with probability $p$ and $y_i = 1$ otherwise; if $x_i \in \{x \; | \; f(x) > 0 \}$ we assign $y_i = 1$ with probability $p$ and $y_i = 0$ otherwise.




For various settings of $p$, we performed $200$ test runs for a setting with $10$ input and $10$ output dimensions. Each test run consisted of sampling one function $f$ and $1000$ data points, out of which $600$ were used for optimizing $\theta$ and $400$ were used as a test set to produce the evaluation results shown in Fig.~\ref{tab:results-log-reg}.


\begin{figure}[t]
\centering
\begin{tabular}{c|c|c}
 p & $\mu$  & $\sigma$ \\ \hline
 1.0 & 0.97 & 0.01  \\
 0.9 & 0.86 & 0.02  \\
 0.8 & 0.76 & 0.02  \\
 0.7 & 0.66 & 0.03  \\
 0.6 & 0.56 & 0.03  \\
 0.5 & 0.50 & 0.03  \\
\end{tabular}%
\caption{Test results for logistic regression. $p$ is the run parameter as described in text, $\mu$ is the average ratio of correct predictions, and $\sigma$ is the corresponding standard deviation.}
\label{tab:results-log-reg}
\end{figure}

As this example shows, logistic regression can be translated to QUBO for execution on quantum hardware. This poses an especially interesting application as the function we considered can also be used to encode the behavior of a single neuron within a neural network. However, multiple Taylor approximations deteriorate the results when we want to extend our approach to larger neural networks. Nonetheless, harvesting the power of quantum computing for the optimization of neural networks remains an important challenge for the field~\cite{gabor2020holy}.


\section{Conclusion}
\label{sec:discussion}



Our goal is to simplify the manual process of formulating and transforming a given problem to QUBO form. To this end we have presented a two-step process to generate a QUBO formulation from a general objective function. First, as described in Sec.~\ref{subsec:step1}, multiple approaches like Taylor series expansion can be used to transform an optimization function into a polynomial. Then, as described in Sec.~\ref{subsec:step2}, a specific approach to transform that polynomial into QUBO form can be used: This method consists of replacing real-valued variables with sums of binary variables and introducing new variables to reduce the amount of variables being multiplied. We have modeled ratio cut (Sec.~\ref{ssec:ratio-cut}) and logistic regression (Sec.~\ref{ssec:logistic-regression}) as QUBO problems, hoping that these examples motivate trying out our approaches and help conquering new problems for quantum computing.


Even though we show that our translations conserve the global optimum of the objective functions (cf. the proof of Theorem~1 in the appendix, e.g.) that does not necessarily mean that other properties of the search space are retained. Future work should thus analyse how the newly generated problem formulations fit various solving mechanisms including (most importantly) QAOA and QA.

Naturally, our list of suggested approaches and methods is not complete and can be expanded in further research. We suspect that more wide-spread interest might bring forth the consolidation of a more concise and closed toolkit, which helps developers transform their problem definitions and might support this process with automated procedures as well.


\begin{appendices}
\section{Proof of Theorem~1}

\setcounter{equation}{0}
\setcounter{theorem}{0}

In this appendix, we provide the proof for Theorem~1.
\begin{theorem}\label{thm:correctness}
Let $P$ be a polynomial and $P'''$ the corresponding QUBO form produced according to the procedure we describe in Section II.B, then it holds that 
\begin{equation}\label{eq:theorem-minima}
\min_{x\in L} P(x_1, ..., x_n) = \min_{z\in \mathbb{B}^{n'}} P'''(z_1, ..., z_{n'})
\end{equation}
where $L = \{(x_1, ..., x_n) \; | \;  x_i \in \Lambda_i\}$ is the set of all possible solution candidates of $P$. Furthermore, there is a function $f: \mathbb{B}^{n'} \to L$  such that
\begin{equation}\label{eq:theorem-function}
\forall z^{*} \in \argmin_{z\in \mathbb{B}^{n'}}  P'''(z_1, ..., z_{n'}) : f(z^{*}) \in \argmin_{x\in L} P(x_1, ..., x_n).
\end{equation}
\end{theorem}

\setcounter{equation}{27}

\begin{proof}

In the following we will show the correctness of Eq.~\ref{eq:theorem-minima} using a proof by cases ($\geq$ and $\leq$), where at the same time case $\leq$ also proves correctness of Eq.~\ref{eq:theorem-function}.

Case $\geq$: Let $x \in L$ be an arbitrary solution within the domain of the minimization problem of $P$. Using the method presented in Section II.B of the main paper, an equivalent binary representation $y_x$ can be constructed, which thus corresponds to a solution of $P'''$, i.e., the problem in QUBO form. $y_x$ fulfills the consistency of the variables (a) by construction and (b) due to known $x$, which is why the constraints introduced in Eq.~9 
are automatically fulfilled and corresponding penalty values (Eq.~10
) do not apply. Thus
\begin{equation}
    P(x_1, ..., x_n) = P'''(y_{x_1}, ..., y_{x_{n'}}) \geq \min_{z\in \mathbb{B}^{n'}} P'''(z_1, ..., z_{n'})
\end{equation}
applies, since the transformation from $P$ to $P'''$ leaves the given minimum unchanged, because only the representation of existing variables is changed and new variables are introduced. In particular, the following holds:
\begin{equation}
\min_{x\in L} P(x_1, ..., x_n) \geq \min_{z\in \mathbb{B}^{n'}} P'''(z_1, ..., z_{n'})
\end{equation}

Case $\leq$: Let $z \in \mathbb{B}^{n'}$ be an arbitrary solution of $P'''$. Two cases are now considered: Either $z$ violates the constraints for maintaining consistency between the variables (see Eq.~9 
or $z$ fulfills these constraints and thus the consistency is given.

Let us in the first case assume that $z$ violates the implicit constraints. Then the penalty term from Eq.~10 
takes a value of at least $1$ and it is also multiplied by weighting factor $A$ (Eq.~11
). Since $A$ is constructed using all the coefficients of polynomial $P'$, it holds $\forall x \in L: A > 2P(x_1, ..., x_n)$ and --- because $P'''$ is composed of $P'$ plus the penalty values --- in particular
\begin{equation}\label{eq:case1-proof}
    P'''(z_1, ..., z_{n'}) > \frac{A}{2} > \min_{x\in L} P(x_1, ..., x_n)
\end{equation}

Let us now assume for the second case that $z$ actually fulfills the constraints for maintaining consistency. Then it is possible to extract the information for restoring a solution $x_z \in L$ directly from $z$. In order to do this, the binary representations of the different variables of $x_z$ must first be read out directly from the binary variables of $z$ and finally be transferred into real numbers. This can be done using a procedure that converts solutions of QUBO problem $P'''$ into solutions of original problem $P$, which exactly reflects the task of function $f$ sought. In particular the following applies to such solutions:
\begin{equation}\label{eq:case2-proof}
    P'''(z_1, ..., z_{n'}) = P(x_{z_1}, ..., x_{z_n}) \geq \min_{x\in L} P(x_1, ..., x_n)
\end{equation}

From Eq. \ref{eq:case1-proof} and Eq. \ref{eq:case2-proof} it now follows
\begin{equation}
\min_{x\in L} P(x_1, ..., x_n) \leq \min_{z\in \mathbb{B}^{n'}} P'''(z_1, ..., z_{n'})
\end{equation}

In summary, we have now shown by means of cases $\leq$ and $\geq$ that the minima of original polynomial $P$ are identical to those of polynomial $P'''$ in QUBO form and thus Eq. \ref{eq:theorem-minima} applies. In addition, we have shown within case $\leq$ that every minimal solution of $P'''$ can be transformed into a minimal solution of $P$ and thus Eq. \ref{eq:theorem-function} holds.
\end{proof}

\end{appendices}

\end{document}